\documentclass{article}
\usepackage[utf8]{inputenc}
\usepackage{authblk}

\usepackage{graphicx}
\usepackage{mathtools}
\usepackage{amsmath}
\usepackage{graphicx}
\graphicspath{ {./images/} }
\usepackage{hyphenat}
\usepackage{latexsym}
\usepackage{stackrel,amssymb}
\usepackage{esint}

\title{Optimization of the Euro-Asian network of gravitational detectors for detecting the radiation of collapsing objects}
\author[1,2,3]{Valentin Rudenko \thanks{Corresponding Author:  valentin.rudenko@gmail.com}}
\author[3]{Svetlana Andrusenko}
\author[1,3]{ Daniil Krichevskiy}
\author[1,3]{Gevorg Manucharyan}
\affil[1]{Sternberg Astronomical Institute, Lomonosov Moscow State University}
\affil[2]{Faculty of Physics, Lomonosov Moscow State University}
\affil[3]{Department of Physics, Bauman Moscow State Technical University}
\date{}

\begin{document}

\maketitle

\begin{abstract}
    A Euro-Asian network of four gravitational-wave (GW) interferometers is considered, taking into account the plan to create such a detector in Novosibirsk. The efficiency of the network is assessed by typical numerical criteria, which also depend on the characteristics of the received signal. In this work, we calculate the optimal orientation of the Novosibirsk detector for the problem of detecting GW radiation accompanying the collapse of the progenitor star with an initial angular momentum. The specificity of the scenario is the presence of the so-called. bar stage deformation, for which the shape of the emitted GW signal is known.
\end{abstract}

\section{Introduction}%вступление

In September 2015, the first direct registration of a gravitational-wave burst from the merger of a relativistic binary, whose components were evaluated as black holes (BH), took place. The detection of this event was carried out using LIGO detectors \cite{Abbott061102}. This was followed by the registration of a big number of events. A qualitative step was the registration by three detectors (including a similar interferometer VIRGO in Europe) GW170814 burst from the merger of BH binary ($M = 30 { M }_{ \odot } $) from the distance of $540 $ Mps \cite{Abbott141101}, which allowed to reduce the localization zone of the source on the celestial sphere by an order of magnitude, up to $\sim 60$ $deg^{2}$. A gravitational wave (GW) signal from neutron stars (NS) merger was registered, coinciding with GRB170817A gamma burst (with $1.7$ $s$ delay) \cite{Abbott161101}.  One of the most recent significant steps is the registration of gravitational waves from the neutron star – black hole binaries coalescence \cite{Abbott_2021}.
All these facts allow to claim confidently real occurrence of a new gravitational-wave channel of astrophysical information and heuristic value of multi-messenger astronomy, i.e. strategy of parallel observation of transients on detectors of different physical nature.

However, there is still no observation of gravitational waves from supernova as well as significant coincidence between LIGO/VIGRO data and neutrino detectors like NOvA \cite{PhysRevD.101.112006} and IceCube \cite{Veske:2021Q6} experiments.  Gravitational waves from core-collapse supernovae can serve as an important source of information about the processes occurring during a given event \cite{Fryer2011}.

In this situation, the problem of creating an optimal terrestrial network of gravitational antennas both in terms of their location, orientation and interconnection  (methods of processing common signals against the background of noises) is of current interest  \cite{gus-ru}. In this context, we continue the discussion of a European-Asian network (EAN) which consists of  four antennas in the northern hemisphere: VIRGO in Italy, KAGRA in Japan, LIGO-India in India and the planned new additional antenna in Novosibirsk. In order to assess the scientific feasibility and eﬀiciency of such a network  the calculation of its main
characteristics was performed in the approach developed in the articles  \cite{Raffai2013,Raffai2015}. In this paper we consider  a gravitational signal from core-collapse supernovae and define the optimal orientation of Novosibirsk detector while the analysis for the case of relativistic binary coalescence was carried out in the previous paper \cite{universe6090140}.
 In order to select the optimal detector orientation angle in Novosibirsk, we perform numerical calculation of  criteria \cite{Raffai2013}. On this basis, a generalized integral efficiency criterion is formed, maximizing which by the angle of orientation of the detector, we find the most effective angle of Novosibirsk detector.

\begin{figure}[h]
\begin{center}
\includegraphics[width=25pc]{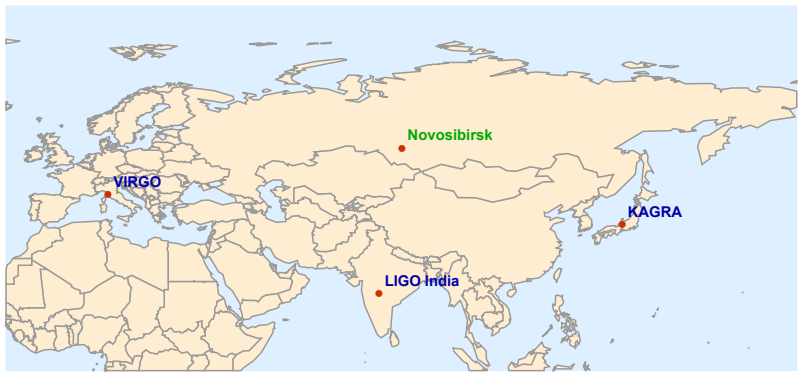}
\end{center}
\caption{\label{Geo}Geographic location of all detectors.}
\end{figure}

Table \ref{Table 1.} shows the coordinates of the detectors in question. The detector orientation angle $ \gamma $ is defined as the angle between the southward direction at the detector location and the  bisector of the angle formed by its arms, measured counterclockwise. Figure \ref{Geo} shows geographic location of all detectors.

\begin{table}[h] %таблица с координатами детекторов
\caption{\label{Table 1.}Detector data; all angles given in degrees.}
\begin{center}
\begin{tabular}{llll}

Detector&Latitude $l$&Longitude $L$&Orientation $\gamma$\\

VIRGO & 43.6 & -10.5 & 206.5\\
KAGRA & 36.4 & -137.3 & 163.3\\
LIGO India & 19.6 & -77.0 & 254.0\\
Novosibirsk & 55.0  & -82.9 & to be defined\\

\end{tabular}
\end{center}
\end{table}

\section{Criteria of a network} %критерии качества

To estimate efficiency of a network of ground based detectors it's necessary to construct power patterns of individual components and the whole network. Let's result the basic information necessary for construction of the pattern.

In the long wavelength approximation (the GW wavelength is much larger than the interferometer arm length $L$) the detector response can be evaluated as
\begin{equation}
h(t)=\frac { \delta L }{ L } ={ F }_{ + }(\theta,\varphi,\psi ){ h }_{ + }(t)+{ F }_{ \times }(\theta,\varphi,\psi ){ h }_{ \times }(t),
\end{equation}
where ${ F }_{ + }(\theta, \varphi, \psi )$, ${ F }_{ \times }(\theta, \varphi, \psi )$ are the antenna pattern functions for the two polarizations, which are functions of the polar angle  $\theta$ and the azimuth angle $\varphi$ of the spherical coordinate system (XY is the detector plane) and the polarization angle of the GW $\psi$.

 Antenna pattern functions have the following form (in the coordinate frame, which basis vectors, coincide with the direction of the detector arms):
\begin{equation}
    { F }_{ + }=\frac { 1 }{ 2 } (1+\cos ^{ 2 }{ \theta } )\cos { 2\varphi } \cos { 2\psi } -\cos { \theta } \sin { 2\varphi } \sin { 2\psi } ,
\end{equation}

\begin{equation}
    { F }_{ \times }=\frac { 1 }{ 2 } (1+\cos ^{ 2 }{ \theta } )\cos { 2\varphi } \sin { 2\psi } +\cos { \theta } \sin { 2\varphi } \cos { 2\psi }.
\end{equation}

\begin{figure}[h]
\begin{center}
\includegraphics[width=20pc]{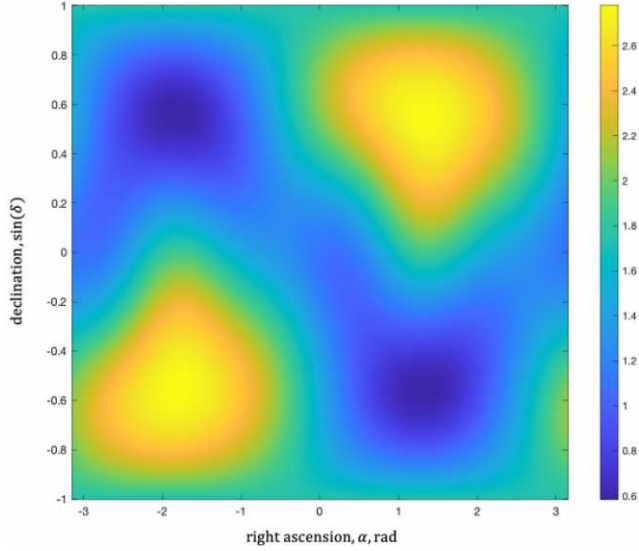}
\end{center}
\caption{\label{EAN} EAN network power pattern.}
\end{figure}

In \cite{Finn2001} it's shown that for a network of $N$ detectors network antenna power pattern ${ P }^{ N }$:

\begin{equation}
    { P }^{ N }=\sum _{ k=1 }^{ 4 }{({ F }_{ +,k }^{ 2 }+{ F }_{ \times,k }^{ 2 }) }.
\end{equation}
Figure \ref{EAN} shows EAN network power pattern.

In order to choose the optimal detector angle in Novosibirsk, we use 3 independent criteria presented in the works \cite{Raffai2013,Raffai2015}. These three conditions form an integral criterion, which is to be maximized by changing the orientation of the Novosibirsk detector, to find the most effective angle.

\subsection{Polarization criterion I}
\unskip
Criterion I characterizes ability of the network to assess the polarization of the received GW. Following \cite{Raffai2013} we define $+$ and $\times$ integral functions for a network of four detectors:
\begin{equation}
{ F }^{ N }=\frac { 1 }{ 2 } \sqrt { { F }_{ 1 }^{ 2 }+{ F }_{ 2 }^{ 2 }+{ F }_{ 3 }^{ 2 }+{ F }_{ 4 }^{ 2 } } ,
\end{equation}
where $N$ - stands for a network function and ${F}_{1}...{F}_{4}$ all either correspond to the $+$ or $\times$ polarization. Obviously ${F }^{ N }$ depend on the polarization angle of $\psi $. 

Calculation of I is carried out in  dominant polarization frame (DPF) \cite{Klimenko2011}.  In DPF for each point on the celestial sphere $(\alpha ; \delta)$ (in equatorial coordinate system point is defined by right ascension $ \alpha \in [-\pi ;\pi ] $ and declination $ \delta \in [-\frac { \pi }{ 2 } ;\frac { \pi }{ 2 }] $) a polarization angle that maximizes the network factor ${ F }_{+}^{ N }$ and minimizes ${ F }_{\times}^{ N }$ is chosen.  Consequently, for this direction $(\alpha ; \delta)$ the condition ${ F }_{+}^{ N } \ge { F }_{\times}^{ N } $ is valid. The condition of approximate equality of factors ${ F }_{\times}^{N}$ and $ { F }_{+}^{N}$ has to be kept, i.e. $\frac {{ F }_{\times}^{ N }  }{ { F }_{+}^{ N } }\approx 1 $. This means that the gravity detector network will be sensitive to both gravity wave polarizations. It follows that a minimum difference of $\left| { F }_{\times}^{ N } - { F }_{+}^{ N }  \right| $ should be sought for all $(\alpha ; \delta)$. This leads to the quantitative formulation of the polarization criterion I \cite{Raffai2013}: 
\begin{equation}
I={ \left(\frac { 1 }{ 4\pi } \oiint { { \left| { F }_{ + }^{ N }(\alpha ; \delta)-{ F }_{ \times }^{ N }(\alpha ; \delta) \right| }^{ 2 }d\Omega } \right) }^{ -1/2 },
\end{equation}
where averaging of $\left| { F }_{\times}^{ N } - { F }_{+}^{ N }  \right| $ over celestial sphere takes place ($d\Omega$ - solid angle). The bigger is $I$ the smaller is the averaged difference $\left| F^{N}_{\times }-F^{N}_{+}\right|  $.

\subsection{Localization criterion D}
Criterion D characterizes the ability of a network to define angular position of a source. In astrometry the problem of a source localization on celestial sphere of a radiation source is solved by a method of triangulation. 
Triangulation is based on the difference in time between the registration of signals by network detectors. The further apart the detectors are, the greater the time delay is. To maximize the source location accuracy on the celestial sphere, the telescopes should be placed as far apart from each other as possible. According to \cite{Raffai2013}, for a network of four detectors $D$ is calculated as the area of the triangle formed by the three detectors in the network, which has the largest area among all possible combinations. If $O$ - the center of the Earth, $A$, $B$, $C$ - points where the detectors are located, the area of the corresponding triangle:
\begin{equation}
S_{ABC}=1/2|[\overrightarrow{AC},\overrightarrow{AB}]|=1/2|[\overrightarrow{OC}-\overrightarrow{OA};\overrightarrow{OB}-\overrightarrow{OA}]|.
\end{equation}

\subsection{Parameters reconstruction criterion R }
Criterion R characterizes the possibility of reconstruction the parameters of the signal of a known analytical form. According to the Maximum likelihood estimation in the additive Gaussian noise background model, the parameters of the received signal are evaluated by the Rao-Cramer bound. The best possible estimates are obtained using the Fisher information matrix $ { \Gamma }_{ \alpha \beta }$  \cite{Kocsis2007} in accordance with the formula
\begin{equation}\label{FisherMatrix}
{ \Gamma }_{ \alpha \beta }=Re\left\{ 4\int _{ { f }_{ min } }^{ { f }_{ max } }{ \frac { \overline { { \partial }_{ \alpha }\tilde { h } (f) } { \partial }_{ \beta }\tilde { h } (f) }{ { S }_{ n }(f) } df } \right\},
\end{equation}
where $\tilde { h } (f)$  is the Fourier image of response of the detector, the line above the Fourier image of response represents the complex conjugate, and ${S }_{ n}(f)$ is the spectral noise density of a single detector. In this paper we assume for simplicity that all detectors have the same noise properties presented in \cite{O1data}.

Rao-Cramer bound  determines the best possible accuracy of parameter $P$ estimation \cite{Kocsis2007}:
\begin{equation}
{ \delta P }^{ 2 }={ \left( { \Gamma }_{N}^{ -1 } \right) }_{PP},
\end{equation}
where ${ \Gamma }_{N}=\sum _{ i=1 }^{ N }{ { \Gamma }_{ i }^{ } } $, i.e. the Fisher information matrix for detector network, is the sum of the corresponding detector matrices constituting the network. The inverse value of the celestial-averaged relative error is a numerical expression of criterion $R$: 
\begin{equation}\label{eq:R}
R={ \left( \frac { 1 }{ 4\pi } \oiint { { \left( \frac { \delta P }{ P } \right) }^{ 2 }d\Omega } \right) }^{ { -1 }/{ 2 } }={ \left< \frac { \delta P }{ P } \right> }^{ -1 },
\end{equation}
Maximization of criterion $ R $ leads to the minimum relative error averaged over the celestial sphere in the estimation of the parameter.

\subsection{ Integral criterion C }
Integral criterion $C$ is used to compare different configurations of a network:
\begin{equation}
C=\sqrt{(\frac{I}{I_{max}})^2+(\frac{D}{D_{max}})^2+(\frac{R}{R_{max}})^2}
\end{equation}
Criteria $ I $, $ D $ and $ R $ together define a three-dimensional space that can be used to define a point which describes a particular configuration. Maximization of $C$ by orientation angle of the detector in Novosibirsk gives the optimal orientation angle and leads to values $I(\gamma^{max}_{Nsk} )=I_{max}, D(\gamma^{max}_{Nsk})= D_{max}$ and $R(\gamma^{max}_{Nsk})=R_{max}$ (from the definition it's clear $C_{max}=1$).

\section{Source} %источник

In this paper we considered EAN efficiency of in registration of gravitational waves from core-collapse supernovae. During the core-collapse there exist many mechanisms of gravitational waves radiation on different stages of the process \cite{PhysRevD.93.042002}. As Fisher matrix approach (\ref{FisherMatrix}) requires analytical form of the signal we have considered gravitational waves from long-lived rotational instabilities of a proto-neutron star. If the key result of this instabilities is bar deformation that the radiation can be simulated by radiation from a rotating cylinder (axis of rotation is a bisector of the cylinder axis) with a Gaussian exponent which is introduced phenomenologically to take into account finiteness of the signal:
\begin{equation}
    h(t)=\sqrt{F^{2}_{+}+F^{2}_{\times }} \frac{GM\omega^{2} L^{2}}{3c^{4}r} (1-3\varepsilon^{2} )e^{-\frac{4(t-t_{0})^{2}}{t^{2}_{0}} }cos(2\omega t+\phi ),
\end{equation}
\begin{equation}
    \phi =\pi +\tan^{-1} \frac{F_{+}}{F_{\times }},
\end{equation}
where $G$ is Newtonian constant of gravitation, $c$ - speed of light, $M$ - mass of the source, $L$ - length of the cylinder, $R$ - radius of the cylinder, $\omega$ - source angular frequency, $t_{0}$ - characteristic signal length and parameter $\varepsilon=\frac{R}{L}$  characterizes the degree of deformation and is used for calculation of criterion $R$ (\ref{eq:R}). Gravitational radiation carries away energy from the system so $\omega$ decreases over time. However for typical values of source parameters \cite{PhysRevD.93.042002} $\omega =2\pi \times 700$ rad/s, $L=20$ km, $R=5$ km, $M=M_{\odot}$ and $t_{0}=1$ s calculation of gravitational waves luminosity via Einstein's formula \cite{landau1980classical}:
\begin{equation}
 L_{GW}=\frac{G}{5c^{5}} \left< \sum^{3}_{j,k=1} \dddot{Q}_{jk} (t-\frac{r}{c} )\dddot{Q}_{jk} (t-\frac{r}{c} )\right>,
\end{equation}
where $Q_{jk}$ - reduced quadrupole moment and dot denotes time derivative, leads to $\frac{\Delta \omega }{\omega } \sim 10^{-7} $, so we consider $\omega$ to be constant.

\section{Numerical results and conclusions } %результаты для коллапса и заключение
\begin{figure}[h]
\begin{center}
\includegraphics[width=25pc]{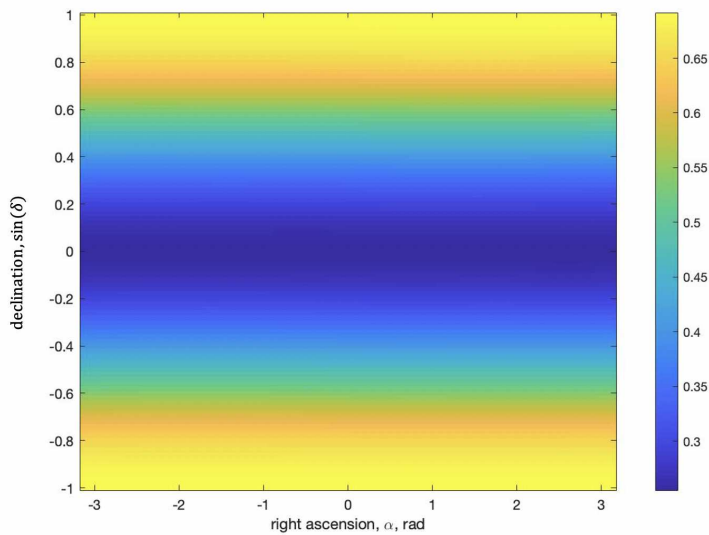}
\end{center}
\caption{\label{avereged}Averaged antenna power pattern for detector in Novosibirsk.}
\end{figure}

In the case of gravitational collapse the strain $h(t)$ is expected to be several orders of magnitude less than for binary coalescence ($h\sim 10^{-22}-10^{-20}$ at $10$ kpc) so we can't assume isotropic distribution of sources over the celestial sphere. Instead we integrate over the Milky Way disk (neglecting the fact that the sensitivity may be sufficient for detection of the signal from several nearby galaxies). Due to rotation of the Earth antenna pattern functions ${ F }_{ + }$ and ${ F }_{ \times }$ depend on time (and consequently $P=P(t)$, $\phi=\phi (t)$) thus we average functions of time over 24 hours which also makes analytical calculation of Fourier image $\tilde { h } (f)$ in (\ref{FisherMatrix}) possible (see Figure \ref{avereged} as an example).

\begin{figure}[h]
\begin{center}
\includegraphics[width=20pc]{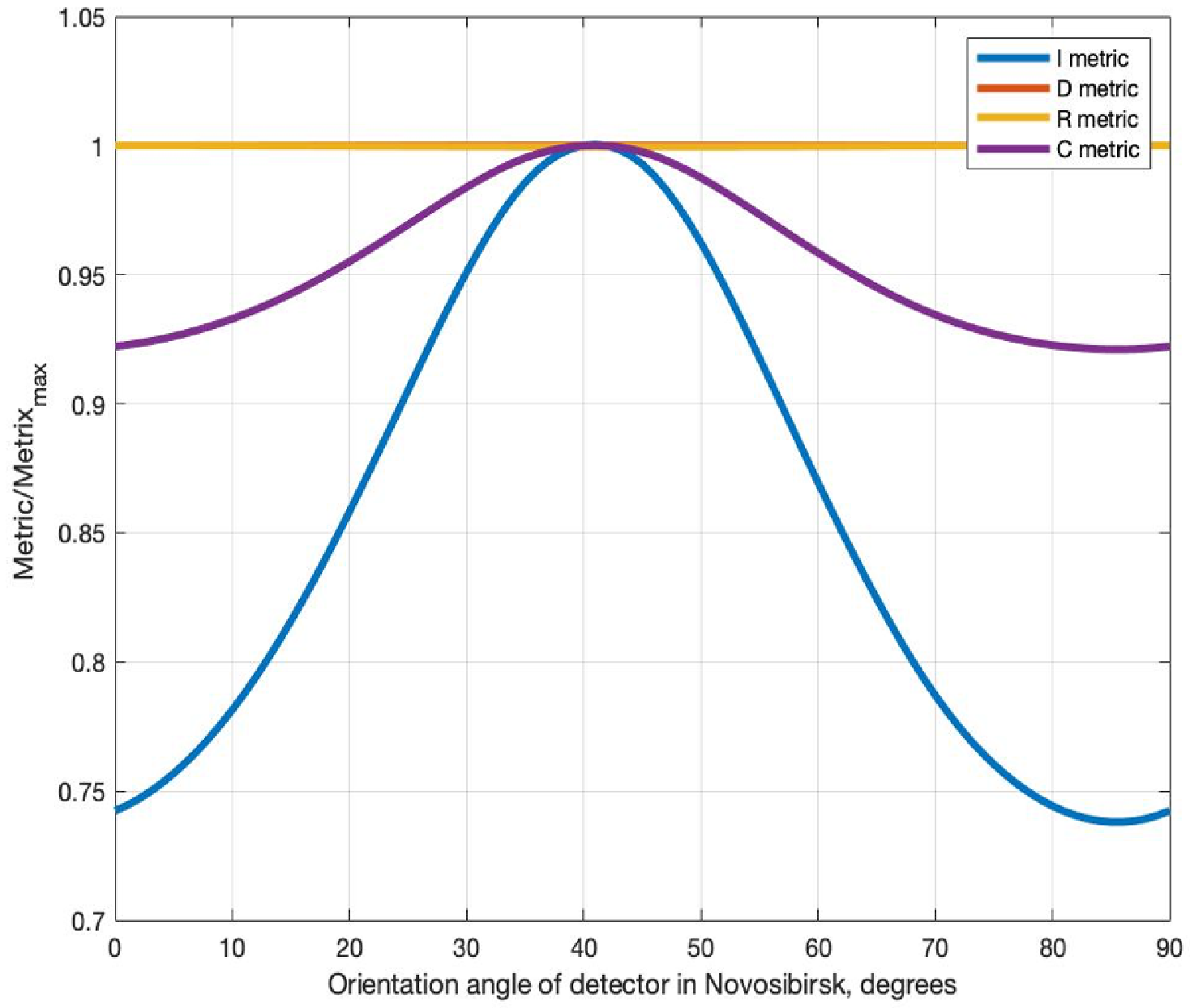}
\end{center}
\caption{\label{AllCriteria}Dependence of all criteria on orientation angle of detector in Novosibirsk.}
\end{figure}

\begin{figure}[h]
\begin{center}
\includegraphics[width=20pc]{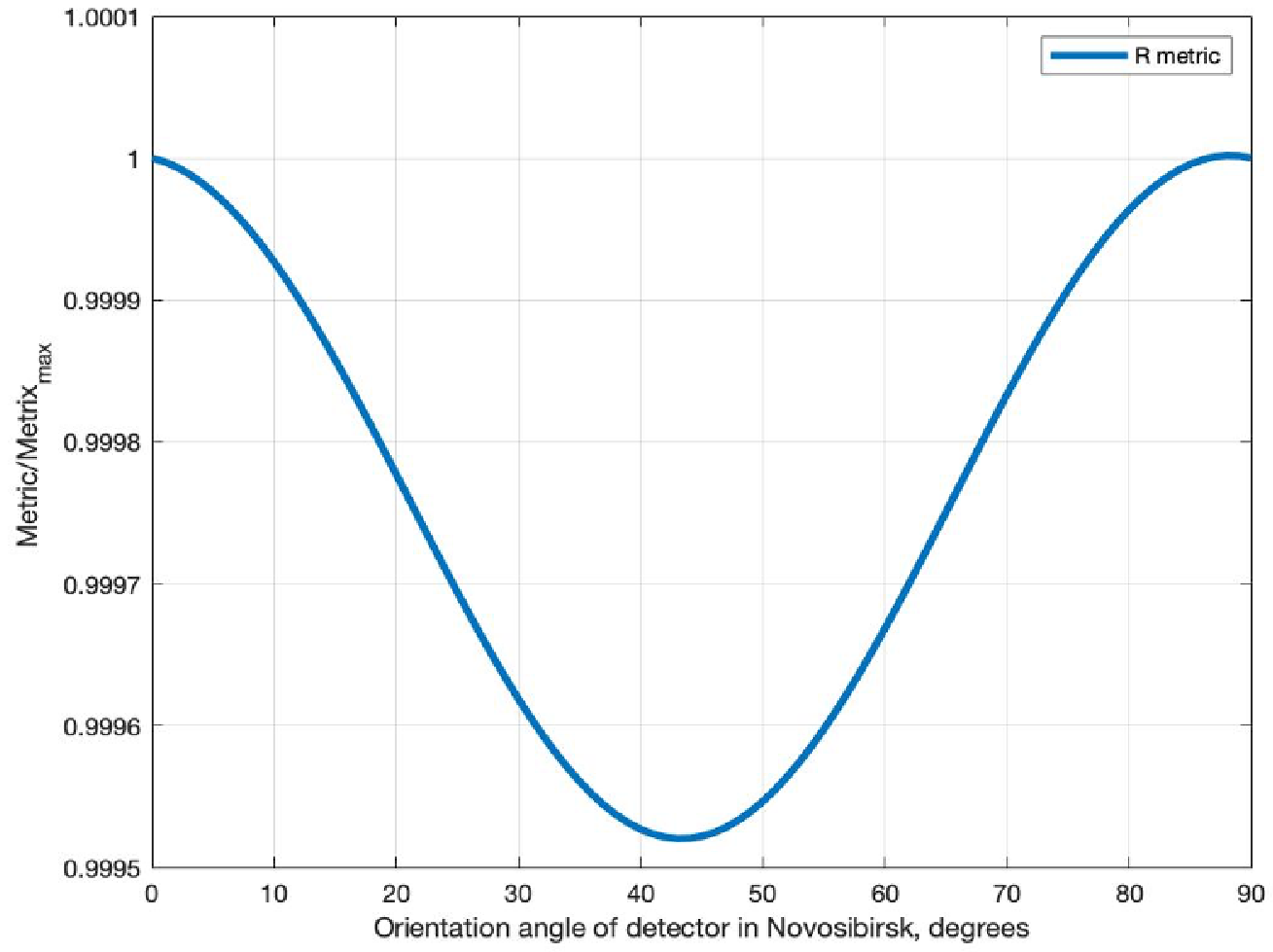}
\end{center}
\caption{\label{R_criterion}Dependence of R criterion on orientation angle of detector in Novosibirsk.}
\end{figure}

\begin{figure}[h]
\begin{center}
\includegraphics[width=20pc]{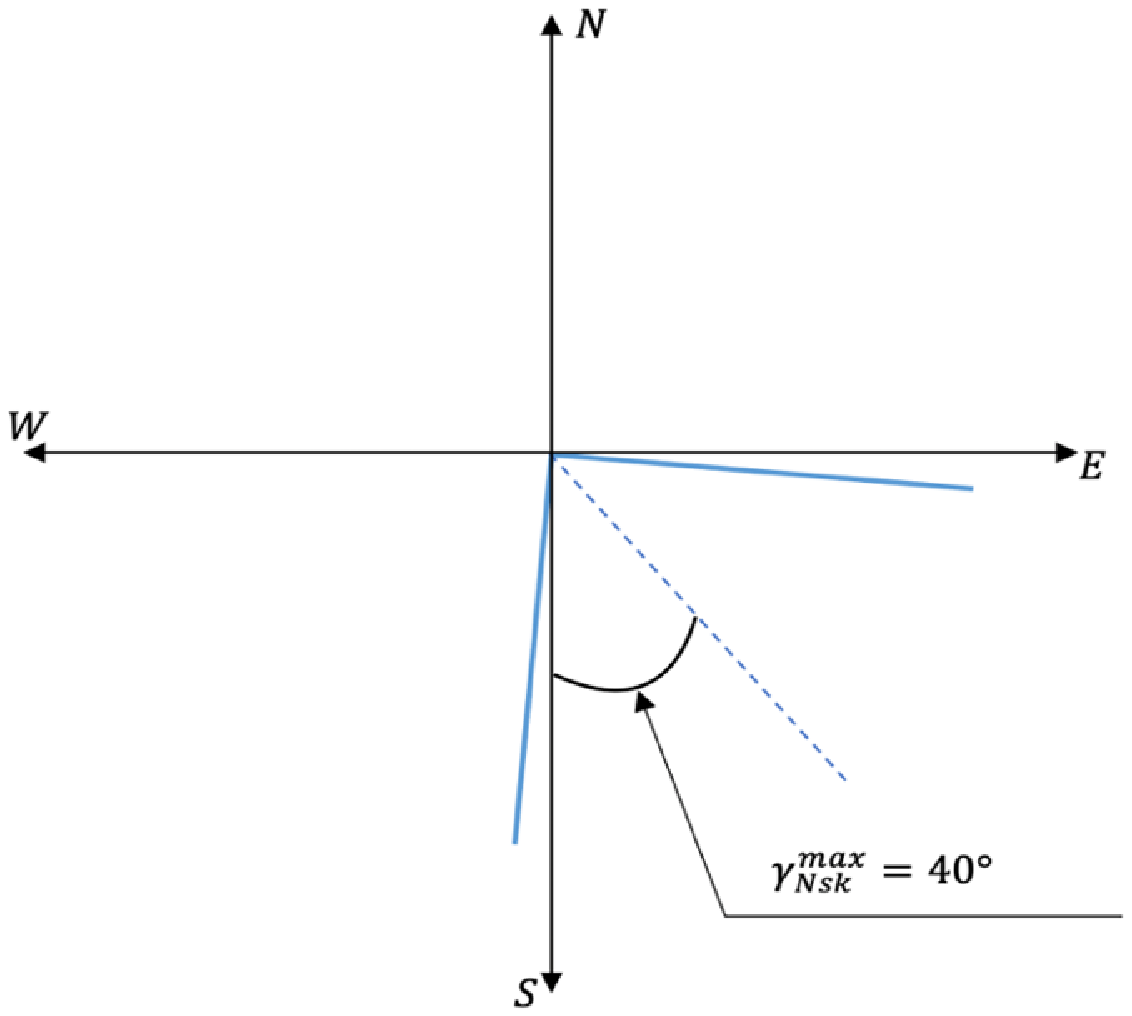}
\end{center}
\caption{\label{orientation}Optimal orientation of the detector in Novosibirsk.}
\end{figure}

The calculation results are shown in Figure \ref{AllCriteria} and Figure \ref{R_criterion}.
Criterion $ D $ obviously  does not depend on the orientation angle of the detector in Novosibirsk. $ I $ is the most sensitive criterion to changes of orientation angle. The maximum value of integral criterion $ C $ is achieved at ${ \gamma  }_{ Nsk }^{max}={ 40 }^{ \circ  } $ (see Figure \ref{orientation}). It's worth noting that the optimal angle differs from the case of binary coalescence - ${ 13 }^{ \circ  }$ \cite{universe6090140}. The final choice depends on the most topical problems at the time of construction of the detector in Novosibirsk. To conclude we notice that the choice of the source is a limitation of this paper, because more significant physics is encrypted in a more complex structure of the signal from the core collapse, but such signals are model-dependent and do not have an analytical form (e.g. \cite{PhysRevD.78.064056}), which does not allow using them within the framework of this approach.

\section*{Acknowledgments}
The authors are grateful to V.I.Pustovoit, A.N.Morozov and V.L.Kauts for their attention to this work and useful discussions.

This work was supported by the grant RFBR 19-29-11010.

\bibliography{References.bib}
\end{document}